\documentclass[preprint,3p]{elsarticle}

\usepackage{amssymb}
\usepackage{amsmath}
\usepackage{color}

\journal{}

\begin{document}

\begin{frontmatter}

\title{
On non-negative solutions to large systems of random linear equations
}

\author[label1]{Stefan Landmann}
\address[label1]{Institute of Physics, Carl von Ossietzky University of Oldenburg,
D-26111 Oldenburg,
Germany}

\ead{stefan.landmann@uol.de}

\author[label1]{Andreas Engel}

\begin{abstract}
Systems of random linear equations may or may not have solutions with all components being non-negative. The question is, e.g., of relevance when the unknowns are concentrations or population sizes. In the present paper we show that if such systems are large the transition between these two possibilities occurs at a sharp value of the ratio between the number of unknowns and the number of equations. We analytically determine this threshold as a function of the statistical properties of the random parameters and show its agreement with numerical simulations. We also make contact with two special cases that have been studied before: the storage problem of a perceptron and the resource competition model of MacArthur.
\end{abstract}


\end{frontmatter}


\section{Introduction}\label{sec1}
Large systems of linear equations show up in many areas of physics. In particular they occur when investigating the stability of stationary states in systems with many degrees of freedom. Quite often, e.g. when considering chemical networks \cite{PoEs,Schnakenberg} or ecological models \cite{may1972will,Schuster} only non-negative solutions are admissible since concentrations have to be positive or zero. 

The interactions in such large systems are as a rule rather complex and not known in detail. On the other hand, macroscopic properties are believed not to depend on all the microscopic specifics. In situations like these models using random parameters have been extremely successful \cite{mezard1987spin}. A paradigmatic case is provided by spin-glass theory \cite{edwards1975theory,sherrington1975solvable}, but also problems from computational complexity \cite{monasson1999determining}, information theory \cite{mezard2009information} and artificial neural networks \cite{engel2001statistical} have been analyzed along these lines. In these approaches emphasis is on self-averaging properties of the systems under consideration. In the thermodynamic limit their probability distributions become sharp and their averages, which are analytically accessible, characterize each typical individual realization of the randomness. 

In the present paper we investigate the existence of non-negative solutions to large sets of random linear equations consisting of $N$ equations for $\alpha N$ unknowns. We are interested in the limit $N\to\infty$ with $\alpha$ staying constant. As we will show there is a sharp transition at a critical threshold $\alpha_c$ from a phase in which typically no non-negative solutions exist to one where such solutions can be found. The value of $\alpha_c$ is self-averaging, i.e., it depends only on the parameters characterizing the distributions of the random variables involved and not on their particular realization. Using the replica-technique we determine $\alpha_c$ analytically and find very good agreement with results from numerical simulations. 

The paper is organized as follows. In the next section we define the problem and map it to the determination of a random volume in $N$-dimensional space. In section~\ref{sec3} we determine the typical value of this volume by a replica calculation. Section~\ref{sec4} discusses the transition and determines the critical line. In section~\ref{sec5} we consider some limiting cases and make contact with related results from the literature. In section~\ref{sec6} we speculate about the behavior of the system away from criticality and finally section~\ref{sec7} contains our conclusions.

\section{Problem and notation}\label{sec2}
We consider an $\alpha N \times N$ random matrix $\hat{a}$ with entries $a_{\mu i},\, \mu=1\dots \alpha N,\, i=1\dots N,$ drawn independently from a Gaussian distribution with mean $A$ and variance $\sigma^2$,
\begin{equation}
 \langle a_{\mu i}\rangle =A,\qquad \langle (a_{\mu i}-A)^2\rangle=\sigma^2.
\end{equation} 
Moreover we choose a random vector $\mathbf{b}$ composed of independent Gaussian components $b_i,\, i=1,\dots, N$, with 
\begin{equation}
 \langle b_i \rangle =B, \qquad \langle (b_i-B)^2\rangle =\frac{\gamma^2}{N}.
\end{equation}  
Here and in the following the brackets $\langle \dots\rangle$ denote the average over the  $a_{\mu i}$ and $b_i$.

We want to know whether the $N$ linear equations 
\begin{equation}
	\hat{a}^T\mathbf{x}=\mathbf{b},
	\label{Equ: Linear Equation System}
\end{equation}
for the $\alpha N$ unknowns $x_\mu$ possess a solution $\mathbf{x}$ with all components being non-negative, $x_\mu\geq 0,\;\mu=1,..,\alpha N$, which we write as $\mathbf{x\geq 0}$. 

The answer to this question clearly depends on the realization of the random numbers $a_{\mu i}$ and $b_i$. In the limit $N\to \infty$ with $\alpha$ and $\gamma$ staying constant, however, there is a critical value $\alpha_c$ of $\alpha$ such that for $\alpha<\alpha_c$ there is no such solution for almost all realization of the random parameters whereas above this threshold typically one can be found. Our aim is to determine $\alpha_c$ as a function of $A,\sigma,B$ and $\gamma$.

\begin{figure}
	\centering
	\includegraphics[width=0.7\linewidth]{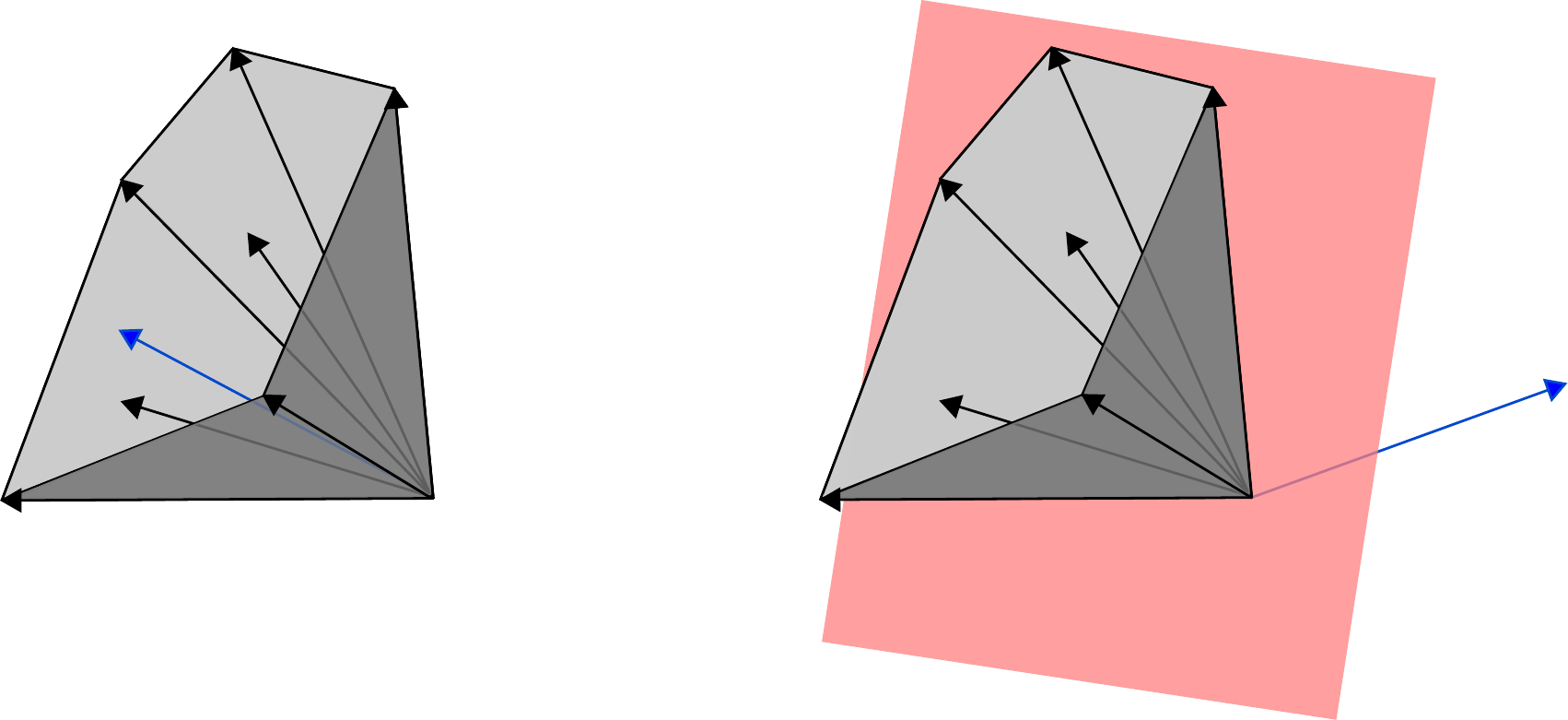}
	\caption{All linear combinations \eqref{eq:defcone} of the column vectors of matrix $\hat{a}$ (black) with non-negative coefficients span a cone in $N$-dimensional space. If the inhomogeneity vector $\mathbf{b}$ (blue) belongs to this cone (left) Eq.~\eqref{Equ: Linear Equation System} has a solution $\mathbf{x}\geq 0$. If $\mathbf{b}$ lies outside the cone (right) no such solution exists. At the same time a hyperplane (red) can be found that separates the cone from  $\mathbf{b}$.}
   \label{fig:farkas}
\end{figure}

It is advantageous to map the problem onto an equivalent one. Let us consider the row vectors $\mathbf{a}_\mu$ of the matrix $\hat{a}$. All their linear combinations 
\begin{equation}\label{eq:defcone}
 c_1 \mathbf{a}_1+c_2 \mathbf{a}_2+\dots + c_{\alpha N} \mathbf{a}_{\alpha N}
\end{equation} 
with non-negative coefficients $c_\mu$ form what is called the non-negative cone of the vectors $\mathbf{a}_\mu$. If $\mathbf{b}$ belongs to this cone, cf. Fig.~\ref{fig:farkas} left, \eqref{Equ: Linear Equation System} has a non-negative solution $\mathbf{x}$. If $\mathbf{b}$ lies outside the cone as shown in  Fig.~\ref{fig:farkas} right no such solution exists. In this case there must be a hyperplane separating $\mathbf{b}$ from the non-negative cone. Denoting the normal of this hyperplane by $\mathbf{y}$ we therefore either have 
\begin{equation}\label{eq:nqw}
   \hat{a}^T\mathbf{x}=\mathbf{b} \,  \, \text{with} \, \,\mathbf{x \geq 0},
\end{equation} 
or there exists a vector $\mathbf{y}$ with 
\begin{equation}\label{eq:y} 
 \hat{a} \mathbf{y} \geq \mathbf{0} \, \, \text{and}  \, \,  \mathbf{b}\cdot\mathbf{y} < 0.
\end{equation} 
This duality is known as Farkas' Lemma \cite{farkas1902theorie}. In what follows we will investigate the problem given by Eq.~\eqref{eq:y}. To this end we determine the typical volume of solutions $\mathbf{y}$ to this problem. If it is positive, \eqref{eq:y} has solutions and consequently \eqref{eq:nqw} has none. If the volume is zero no separating hyperplane exists, $\mathbf{b}$ is part of the non-negative cone and \eqref{eq:nqw} can be solved. The transition between both cases occurs when the volume shrinks to zero. 

\section{Typical volume of solutions to the dual problem}\label{sec3}
\begin{figure}
	\centering
	\includegraphics[width=0.3\linewidth]{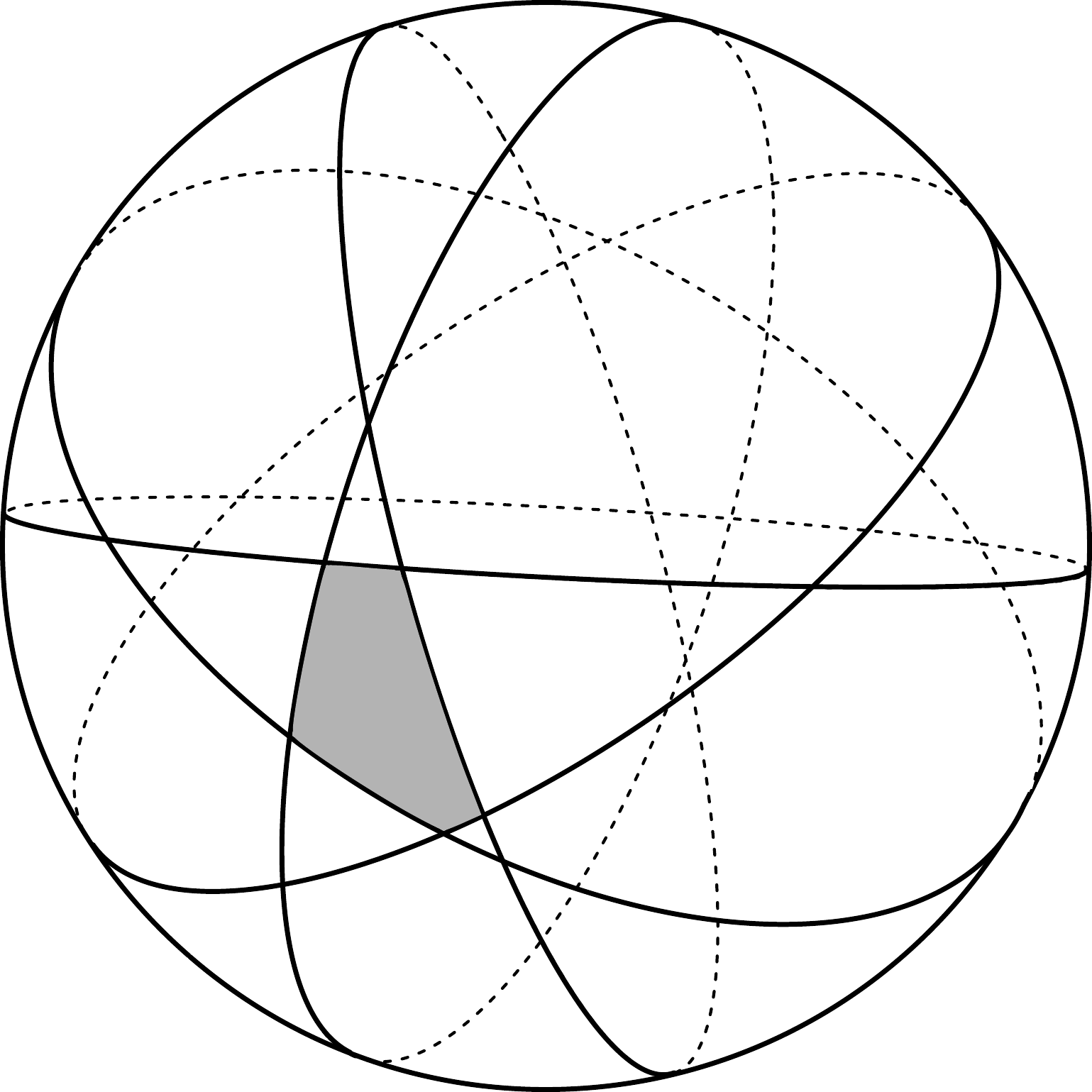}
	\caption{The solution space of the dual problem~\eqref{eq:y} (shaded) is part of the $N$-sphere defined by the spherical constraint~\eqref{eq:sphconstr} and cut by hyperplanes representing the inequalities in~\eqref{eq:y}.}
	\label{fig:sphere}
\end{figure}

With each solution $\mathbf{y}$ to \eqref{eq:y} also $\lambda \mathbf{y}$ with positive $\lambda$ is a solution. In order to eliminate this trivial degeneracy it is convenient to impose the spherical constraint 
\begin{equation}\label{eq:sphconstr}
 \|\mathbf{y}\|^2=\sum_i y_i^2=N
\end{equation} 
on the solution vectors $\mathbf{y}$. Our central quantity of interest is the fractional volume 
\begin{align}
 \Omega (\hat{a},\mathbf{b}):=
	\frac{\int^\infty_{-\infty} \prod_i dy_i\, \delta \left(\sum_i y_i^2-N\right)
		\Theta \left(-\frac{1}{\sqrt{N}}\sum_i  b_i y_i \right)
		\prod_\mu\Theta \left( \frac{1}{\sqrt{N}} \sum_i a_{\mu i} y_i \right)  }{\int^\infty_{-\infty} \prod_i dy_i \,\delta(\sum_i y_i^2-N)}.
	\label{Equ: Volume of solutions}
\end{align}
It comprises all normalized vectors $\mathbf{y}$ obeying the inequalities \eqref{eq:y}, cf. Fig.~\ref{fig:sphere}. In \eqref{Equ: Volume of solutions} $\Theta(x)$ denotes the Heaviside function 
\begin{equation}
 \Theta(x)=\begin{cases} 1 & x\geq 0\\ 0 & x<0 \end{cases}.
\end{equation} 

The expression for the fractional volume is rather similar to that for the version space in neural network models \cite{gardner1988space} and can be analyzed along similar lines. In particular, since $\Omega$ involves the product of many independent random terms its logarithm is assumed to be self-averaging, i.e. in the large $N$ limit the typical value of $\Omega$ is given by 
\begin{equation}
 \Omega_\mathrm{typ}\simeq \exp\left(\langle \log \Omega\rangle\right)
\end{equation} 
rather than by $\langle \Omega \rangle$. Therefore, in order to characterize the solution space of \eqref{eq:y} we need to determine the averaged intensive entropy
\begin{equation}\label{defS}
	S(\alpha,A,B,\sigma,\gamma):=\lim_{N\to\infty}\frac{1}{N}
	\langle\log \Omega(\hat{a},\mathbf{b})\rangle.
\end{equation}
This may be done using the replica trick~\cite{mezard1987spin} which is based on the identity
\begin{equation}\label{replica}
	\left< \log  \Omega \right>=\lim_{n\rightarrow 0}\frac{ \left<\Omega^n\right>-1}{n}.
\end{equation}
$\langle\Omega^n\rangle$ is first determined for $n\in \mathbb{N}$ and the result needs to be continued in a meaningful way to real $n$ in order to perform the limit $n\to 0$. For $n\in \mathbb{N}$ we have
\begin{align}
	\Omega^n & (\hat{a},\mathbf{b})= \nonumber
	\\
	&\int^\infty_{-\infty} \prod_{i,a} \frac{dy^a_i}{\sqrt{2 \pi e}} \prod_a \delta \left(\sum_i (y^a_i)^2-N \right)\prod_{\mu,a} \Theta \left( \frac{1}{\sqrt{N}} \sum_i a_{\mu i} y^a_i \right)\prod_a \Theta \left(-\frac{1}{\sqrt{N}}\sum_i b_i y^a_i \right),
	\label{Equ:replica Volume of solutions}
\end{align}
where the replica index $a$ runs from 1 to $n$ and the denominator $\sqrt{2 \pi e}$ accounts for the normalization \eqref{eq:sphconstr}. Using standard techniques \cite{engel2001statistical} we replace the $\delta$- and $\Theta$-functions by their integral representations:
\begin{align*}
&\prod_a \delta\left(\sum_i (y^a_i)^2-N \right)=\int \prod_a \frac{dE^a}{4 \pi} 
\text{exp} \left(\frac{i}{2}\sum_a E^a(\sum_i (y^a_i)^2-N)\right),
\\
&\prod_a \Theta \left(-\frac{1}{\sqrt{N}}\sum_i b_i y^a_i \right)=\int^\infty_0\prod_a d\eta^a \int \prod_a \frac{d\hat{\eta}^a}{2 \pi/N }\text{exp}\left( i N  \sum_a \hat{\eta}^a\Big(\eta^a+\frac{1}{\sqrt{N}}b_i\sum_i y^a_i\Big)\right),
\\
&\prod_{\mu,a} \Theta \left( \frac{1}{\sqrt{N}} \sum_i a_{\mu i} y^a_i \right)=\prod_\mu \int^\infty_0 \prod_a d\vartheta^a_\mu \int \prod_a \frac{d\hat{\vartheta}^a_\mu}{2 \pi} \text{exp} \left(i\sum_a \hat{\vartheta}^a_\mu\Big(\vartheta^a_\mu-\frac{1}{\sqrt{N}}\sum_ia_{\mu i}y^a_i\Big) \right),
\end{align*}
and perform the averages over $a_{\mu i}$ and $b_i$ 
\begin{align}
	\prod_{i,\mu} \left< \text{exp} \left(-\frac{i}{\sqrt{N}}\sum_{a} a_{\mu i} \hat{\vartheta}^a_\mu  y^a_i\right)\right> &=
	\prod_{i,\mu} \text{exp}\left\{-\frac{i}{\sqrt{N}} A\sum_a \hat{\vartheta}^a_\mu y^a_i- \frac{\sigma^2}{2N} \left(\sum_a \hat{\vartheta}^a_\mu y^a_i\right)^2  \right\},
	\label{eq: ave a}
	\\
	\prod_i	\left<\text{exp} \left(i\sqrt{N} b_i\sum_a \hat{\eta}^a y^a_i\right) \right>&=\prod_i \text{exp} \left\{i \sqrt{N}B\sum_a \hat{\eta}^ay^a_i -\frac{\gamma^2}{2}\left(\sum_a \hat{\eta}^a y^a_i \right)^2 \right\}.
	\label{eq: ave b}
\end{align}
The integrals over the $y_i$, the auxiliary variables  $\vartheta^a_\mu,\,\eta^a$ and their conjugates $\hat{\vartheta}^a_\mu,\,\hat{\eta}^a$ can be decoupled by introducing the order parameters
\begin{equation}\label{eq:defop}
	m^a=\frac{1}{\sqrt{N}}\sum_i y^a_i\qquad\mathrm{and}\qquad 
	q^{ab}=\frac{1}{N}\sum_i y^a_i y^b_i\quad\mathrm{for}\; a<b.
\end{equation} 
The expression for the $n$-th power of the fractional volume then acquires the form
	\begin{align}\nonumber
	\left< \Omega^n \right>=\int \prod_a &\frac{dm^a d\hat{m}^a}{2 \pi /\sqrt{N}}
	\int \prod_{a<b} \frac{dq^{ab} d\hat{q}^{ab}}{2 \pi /N}\int \prod_a \frac{dE^a}{4 \pi} 
	\int^\infty_0\prod_a d\eta^a \int \prod_a \frac{d\hat{\eta}^a}{2 \pi/N}\\\nonumber
	& \quad\text{exp} 
	\Big( N\Big[\frac{i}{\sqrt{N}}\sum_a m^a\Hat{m}^a+i\sum_{a<b}q^{ab}\hat{q}^{ab}-\frac{i}{2}\sum_a E^a 
	+i \sum_a \hat{\eta}^a \eta^a+ i B  \sum_a \hat{\eta}^a m^a\\
	&  \quad -\frac{\gamma^2}{2}\sum_a (\hat{\eta}^a)^2-\gamma^2 \sum_{a<b} \hat{\eta}^a\hat{\eta}^b q^{ab} +\alpha G_E(m^a,q^{ab})+G_S(E^a,\hat{m}^a,\hat{q}^{ab})\Big]\Big)\label{Omn},
	\end{align}
with the auxiliary functions
	\begin{align}\label{defGE}
	G_E(m^a,q^{ab})=& \nonumber
	\log \int^\infty_0 \prod_a d\vartheta^a \int \prod_a \frac{d\hat{\vartheta}^a}{2 \pi}\, 
	\\
	&\times \quad \text{exp} \Big(i\sum_a \hat{\vartheta}^a \vartheta^a-i A \sum_a \hat{\vartheta}^a m^a -\frac{\sigma^2}{2}\sum_a (\hat{\vartheta}^a)^2-\sigma^2 \sum_{a<b} \hat{\vartheta}^a \hat{\vartheta}^b q^{ab} \Big),
	\end{align} 
and 
	\begin{equation}\label{defGS}
	G_S(E^a,\hat{m}^a,\hat{q}^{ab})=\log \int \prod_a \frac{dy^a}{\sqrt{2 \pi e}}\, \text{exp} 
	\Big(\frac{i}{2}\sum_aE^a (y^a)^2-i\sum_a \Hat{m}^ay^a-i\sum_{a<b}\Hat{q}^{ab}y^a y^b \Big).
	\end{equation} 
To determine the entropy \eqref{defS} we only need the asymptotics of this expression for ${N\to\infty}$ so that the integrals in \eqref{Omn} may be calculated by the saddle-point method. The term $\frac{1}{\sqrt{N}}\sum_a m^a \hat{m}^a$ can be neglected in this limit and will be dropped. 

The solution space of Eq.~\eqref{eq:y} is connected, we therefore assume a replica-symmetric saddle-point and use the ans\"atze
\begin{align}\label{eq.rs1}
	m^a=& \, m, \quad  i\hat{m}^a=-\hat{m}, \quad iE^a=-E,\quad 
	\eta^a=\eta,\quad i\hat{\eta}^a=\hat{\eta} &\forall a,
	\\\label{eq.rs2}
	q^{ab}=& \, q, \, \, \, \, \, \,  i\hat{q}^{ab}=-\hat{q} &\forall a\neq b.
\end{align}
Keeping in mind that for the final limit $n\to 0$  only terms up to order $n$ are needed the expressions for $G_E$ and $G_S$ simplify considerably. We find
	\begin{align}
	G_E(m,q)=n \int Dt\, \log H \Big(\frac{\sqrt{q}\,t-\kappa}{\sqrt{1-q}}\Big) +O(n^2).
	\end{align}
Here the Gaussian measure $Dt:=dt/\sqrt{2\pi}\,e^{-t^2/2}$ was introduced  and the abbreviations 
	\begin{equation}
	H(x):=\int_x^\infty Dt\qquad\mathrm{and}\qquad\kappa:=\frac{m A}{\sigma}
	\end{equation} 
were used. Similar manipulations yield for $G_S$
	\begin{align}
	G_S(E,\hat{m},\hat{q})=-\frac{n}{2}\Big(1+\log(E+\hat{q})\Big) + \frac{n}{2}\frac{\hat{m}^2+\hat{q}}{E+\hat{q}}+O(n^2).
	\end{align}
Also the other terms in \eqref{Omn} can be simplified using \eqref{eq.rs1} and \eqref{eq.rs2}. As a consequence the saddle-point equations for the order parameters $E,\hat{m},\hat{q},\eta$ become algebraic and these parameters can be eliminated. Introducing 
\begin{equation}\label{eq:defzeta}
 \zeta:=\left( \frac{\sigma B}{\gamma A} \right)^2
\end{equation} 
we finally get 
\begin{align}\label{Sres}
	S(\alpha,\zeta)=\mathrm{extr}_{q,\kappa}\Big[\frac{1}{2}\log(1-q)+\frac{q}{2(1-q)}-
          \frac{\zeta}{2}\frac{\kappa^2}{1-q}
	+\alpha\int Dt\, \log H \Big(\frac{\sqrt{q}\,t-\kappa}{\sqrt{1-q}}\Big)\Big].
\end{align} 
The saddle-point values of $q$ and $\kappa$ have to be determined numerically. Note that the final expression~\eqref{Sres} for the averaged entropy depends on the parameters $A, \sigma, B$ and $\gamma$ characterizing the distributions of $a_{\mu i}$ and $b_i$ only through the combination $\zeta$ 
that gives the ratio between the relative variances of $a_{\mu i}$ and $b_i$. 

\begin{figure}
\centering
    \includegraphics[width=0.6\linewidth]{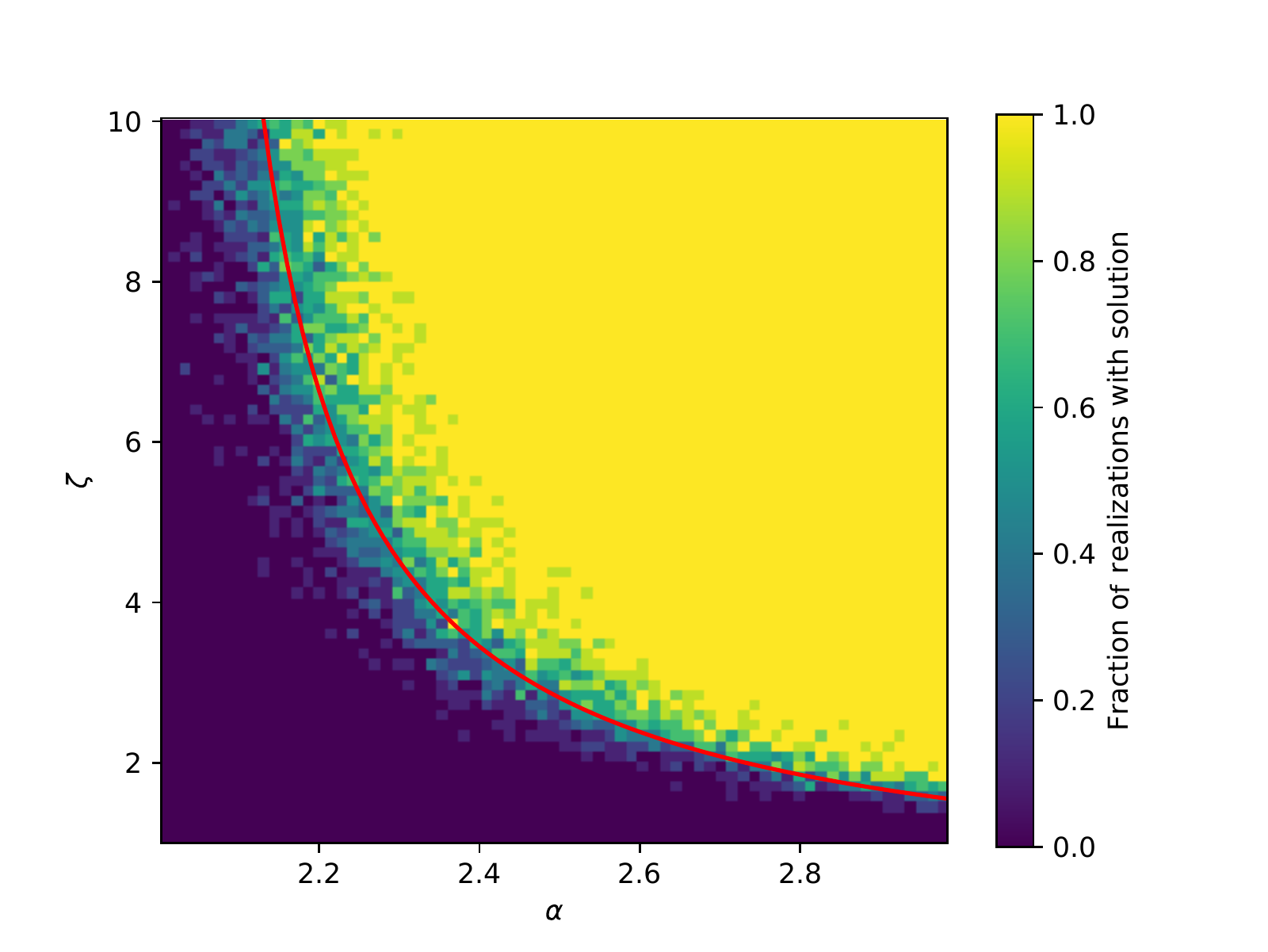}
    \caption{Comparison of the analytical prediction (red line) Eqs.~\eqref{h4} and \eqref{res2} for the phase-transition with simulation results. Each dot shows the fraction of realizations of Eq.~(\ref{Equ: Linear Equation System}) for which a non-negative solution was found. System size is $N=750$, each data point is averaged over 10 realizations.}
    \label{fig:Figure 1}
\end{figure}

\section{The transition point}\label{sec4}
At the transition point $\alpha_c$ the typical volume $\Omega_\mathrm{typ}$ becomes zero, i.e. the average entropy tends to minus infinity. It is possible to extract this point from the numerical extremalization in \eqref{Sres}. However, the analysis can be simplified by observing that with $\Omega$ shrinking to zero the typical overlap $q$ between two different solutions $\mathbf{y}$ as defined in \eqref{eq:defop} tends to one. 
To leading order in $1/(1-q)$ we get
\begin{equation}
	\log H \Big(\frac{\sqrt{q}\,t-\kappa}{\sqrt{1-q}}\Big)\sim
	\begin{cases}
	-\frac{(t-\kappa)^2}{2(1-q)} & \text{if}\quad t>\kappa,\\
	0              & \text{otherwise,}
	\end{cases}
\end{equation} 
implying
\begin{equation}
	\int Dt\, \log H \Big(\frac{\sqrt{q}\,t-\kappa}{\sqrt{1-q}}\Big)
	\sim -\frac{1}{2(1-q)}\int_\kappa^\infty Dt\,(t-\kappa)^2=:-\frac{1}{2(1-q)}\, I(\kappa).
\end{equation} 
Keeping only the most divergent terms near the transition we thus find 
\begin{equation}
	S(\alpha_c,\zeta)\sim\mathrm{extr}_{q,\kappa}\left[\frac{1}{2(1-q)}-\frac{\zeta}{2}\frac{\kappa^2}{(1-q)}-\frac{\alpha_c}{2(1-q)} I(\kappa)\right].
	\end{equation} 
The saddle-point equation with respect to $q$ gives
	\begin{equation}\label{h4}
	1-\zeta \kappa^2=\alpha_c I(\kappa),
	\end{equation} 
whereas the one with respect to $\kappa$ results in 
	\begin{equation}
	-\zeta\kappa=\frac{\alpha_c}{2} \frac{d I}{d\kappa}
	=\frac{\alpha_c}{\kappa}\big(I(\kappa)-H(\kappa)\big).
	\end{equation} 
Together with \eqref{h4} this yields
	\begin{equation}\label{res2}
	\alpha_c H(\kappa)=1.
	\end{equation} 
Eqs.~\eqref{h4} and \eqref{res2} give a parametric description of the transition line $\alpha_c(\zeta)$. It is shown in Fig.~\ref{fig:Figure 1} together with results from numerical solutions of Eq.~(\ref{Equ: Linear Equation System}). There is good agreement between the analytical prediction and simulation results.

\section{Special cases}\label{sec5}

The basic problems \eqref{eq:nqw} and \eqref{eq:y} respectively are rather general. Some special cases have been studied previously in different settings. 

\subsection{Relation to the storage problem of a perceptron}
From the numerical analysis of \eqref{Sres} one finds that for large $\zeta$ the saddle-point value of $\kappa$ tends to zero such that also $\zeta \kappa^2 \to 0$. In this limit we hence find 
\begin{align}\label{Sperc}
	S(\alpha,\zeta\to \infty)=\mathrm{extr}_{q}\Big[\frac{1}{2}\log(1-q)+\frac{q}{2(1-q)}
         +\alpha\int Dt\, \log H \Big(\frac{\sqrt{q}\,t}{\sqrt{1-q}}\Big)\Big]
\end{align} 
which coincides with Gardner's famous expression describing the storage problem of a perceptron \cite{gardner1988space}. In this problem one is given $\alpha N$ random input patterns $\boldsymbol{\xi}_\mu$ and corresponding outputs $\lambda_\mu$ and has to find a synaptic vector $\mathbf{J} \in \mathbb{R}^N$ such that 
\begin{equation}
  \frac{1}{\sqrt{N}}\lambda_\mu \mathbf{J} \boldsymbol{\xi}_\mu \geq 0
\end{equation}
for all $\mu=1, \dots, \alpha N$. For more details see \cite{engel2001statistical}.

\begin{figure}
\centering
    \includegraphics[width=0.6\linewidth]{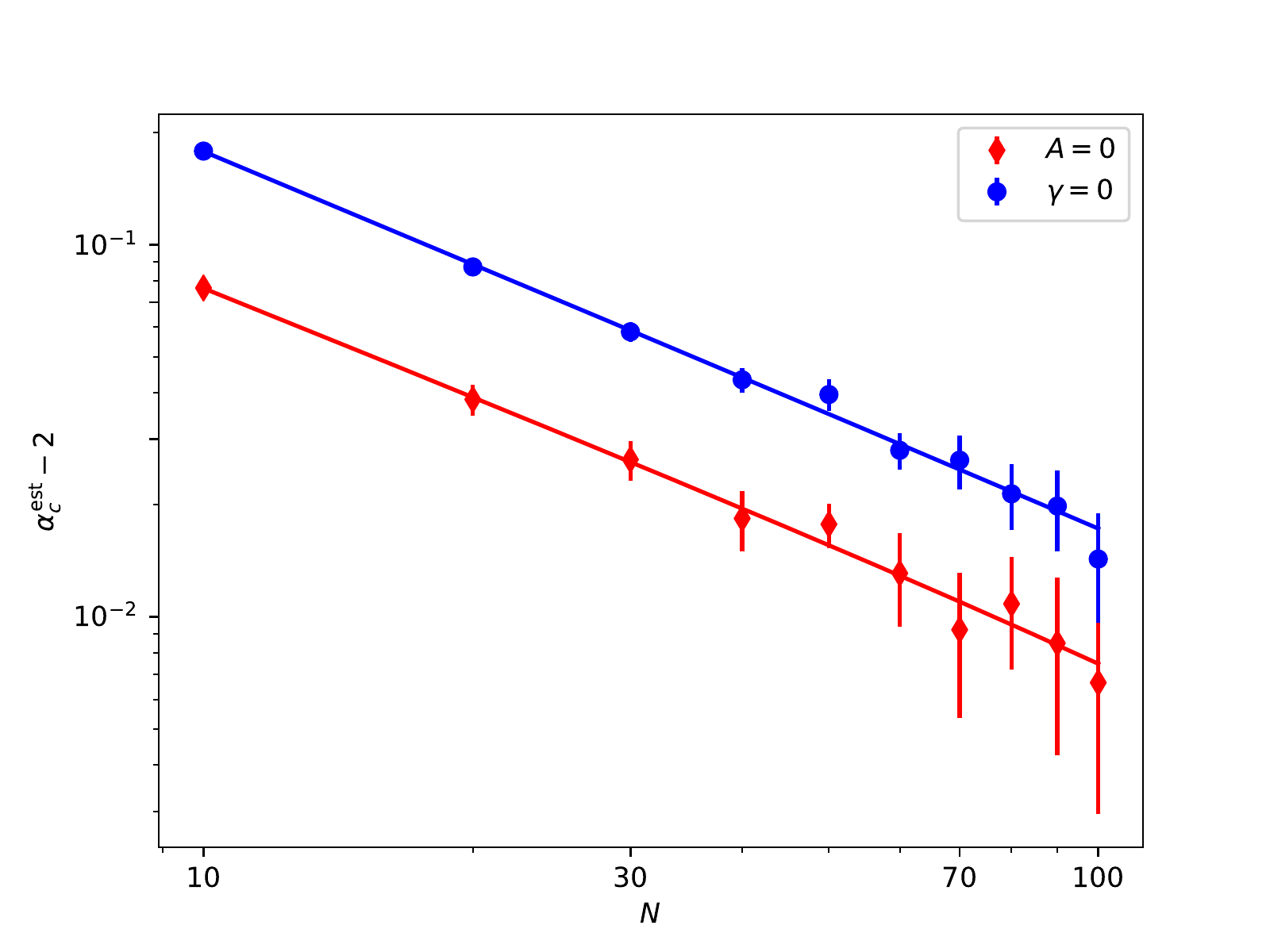}
        \caption{Finite size scaling of the critical $\alpha_c$ for $\zeta=0$, i.e. in the case of equivalence with the storage problem of a perceptron. For $\gamma=0$ (blue,dots) we obtain $\lim_{N \rightarrow  \infty} \alpha^{\mathrm{est}}_c=2.001 \pm 0.004$, and for $A=0$ (red, diamonds) $\lim_{N \rightarrow \infty} \alpha^{\mathrm{est}}_c=2.001 \pm 0.002$ confirming the analytical results. The values of $\alpha^{\text{est}}_c$ were determined as follows: For each system size $N$ and over some range of $\alpha$ values the average number of realizations with non-negative solutions was determined. Then a sigmoidal curve was fitted to this data and $\alpha^{\text{est}}_c$ was chosen as the value where this curve is equal to 0.5.}
    \label{fig:Figure 2}
\end{figure}

There are two situations that make the equivalence between this storage problem for the perceptron and our dual problem \eqref{eq:y} for $\zeta\to\infty$ particularly clear. Firstly, let us consider the case $\gamma=0$ such that $b_i=B$ for all $i$. We then decompose $\Hat{a}=\hat{A}+\delta \, \Hat{a}$ where the entries of $\hat{A}$ are constant $A_{\mu i}=A$ while the entries of $\delta\Hat{a}$ are i.i.d. random gaussian variables with zero mean and variance $\sigma^2$. Then, \eqref{eq:y} requires  
\begin{align} \label{eq:nq}
	 \, \,  \delta\hat{a} \, \mathbf{y} \geq -A\sum_i y_i, \, \, \text{and}  \, \,  \sum_i y_i < 0.
\end{align}
The condition $\sum_i y_i<0$ restricts all $\mathbf{y}$ to lie on the 'negative' half of the $N$-sphere. From high-dimensional geometry it is known that for $N \rightarrow \infty$ the volume spanned by the vectors $\mathbf{y}$ on the $N$-sphere is tightly concentrated around the equator. Therefore, it may be assumed that $\sum_i y_i=0^-$ without reducing the fractional volume~\footnote{This is substantiated by the fact that the saddle-point value of $m$ as defined in \eqref{eq:defop} is zero in this case.}. Effectively the conditions therefore become
\begin{align} \label{eq:nq}
	 \, \,  \delta\hat{a} \, \mathbf{y} \geq \mathbf{0}, \, \, \text{and}  \, \,  \sum_i y_i < 0.
\end{align}
In this form they are equivalent to the storage problem for $\alpha N+1$ patterns. For $N\to\infty$ the influence of the additional pattern can be neglected. The critical value of $\alpha$ in the storage problem is known to be $\alpha_c=2$~\cite{gardner1988space}. Figure~\ref{fig:Figure 2} shows results from a finite size analysis of $\alpha_c$ for the system (\ref{Equ: Linear Equation System}) with $\gamma=0$. For $N\rightarrow\infty$ we obtain $\alpha_c=2.001 \pm 0.004$ corroborating the equivalence between the two problems in the large $N$ limit.

Secondly, our analysis reduces to the storage problem of the perceptron if $A=0$. In this case Eq.\eqref{eq:y} acquires the form
\begin{align} \label{eq:n}
	 \, \,  \delta\hat{a} \, \mathbf{y} \geq \mathbf{0}, \, \, \text{and}  \, \,  B\sum_i y_i+\delta\mathbf{b}\cdot\mathbf{y} < 0.
\end{align}
Clearly, the first condition is equivalent to the storage problem for $\alpha N$ patterns. To show that the second one does not reduce the fractional volume $\Omega$ for large $N$ we split it into $B\sum_i y_i<0$ and $\delta\mathbf{b}\cdot\mathbf{y} < 0$ and require these two constraints separately. Even this more severe restriction is only equivalent to two additional patterns, $\alpha N \to \alpha N +2$, which in the limit $N\to\infty$ is again irrelevant. Figure~\ref{fig:Figure 2} shows that also in this case the finite size analysis of $\alpha_c$ is in perfect agreement with the result from the storage problem of the perceptron.

\subsection{MacArthur model of resource competition}

A classical model to describe biodiversity in ecological systems considers a number $\alpha N$ of species that compete for a number $N$ of different resources \cite{MacArthur}. The species differ in their ability to live on the various resources which is described by a matrix of metabolic strategies~$\hat{a}$. The elements $a_{\mu i}$ of $\hat{a}$ characterize the ease with which species $\mu$ may consume resource~$i$. Resources are provided by certain fixed influxes $b_i$ from the environment. For sufficiently large $\alpha$ there is a stationary state of the system with species concentrations $x_\mu$ in which the overall consumption of each resource $i$ is exactly balanced by its influx:
\begin{equation}\label{MAstationary}
 \sum_\mu a_{\mu i} x_\mu=b_i,\quad i=1,\dots, N.
\end{equation} 
Since moreover all $x_\mu$ must be either positive (if the species survives) or zero (if the species got extinct) the vector $\mathbf{x}$ has to fulfill Eq.~\eqref{eq:nqw}. 

The MacArthur model has been studied extensively for systems with only a few species and correspondingly few resources. On the other hand, realistic biological systems may involve hundreds of species and resources. In a very interesting recent paper \cite{TM17} a random version of MacArthur's model was analyzed in the limit $N\to \infty$. To this end the elements of the matrix of metabolic strategies were chosen independently of each other as either one with probability $p$ or zero with probability $1-p$. Large values of $p$ hence describe communities of universalists that can live from many different resources, small values of $p$ stand for systems consisting mostly of specialists. The resource influx was modelled as a Gaussian random vector with elements of average one and variance $\Delta^2/N$. Interestingly, in this variant of the MacArthur model a transition was found from a so-called vulnerable phase at low potential biodiversity (small $\alpha$) to a so-called shielded phase at large $\alpha$. In the vulnerable phase the species are unable to use all available resources properly, Eq.~\eqref{MAstationary} cannot be fulfilled, less than $N$ species survive, and variations in the resource influxes have a direct impact on each species. Contrarily, in the shielded phase the species form a collective field that guards them from changes in the external conditions, the maximal possible number $N$ of species survives \cite{armstrong1980competitive}, and Eq.~\eqref{MAstationary} is satisfied. This phase is a striking example for the emergence of cooperation in a purely competitive system \cite{Axelrod}. The transition found in \cite{TM17} is equivalent to the one discussed in the present paper for $A=p$, $\sigma^2=p(1-p)$, $B=1$ and $\gamma=\Delta$. For more details see \cite{landmann2018systems}.

\section{Making contact away from criticality}\label{sec6}

\begin{figure}
\centering
    \includegraphics[width=0.6\linewidth]{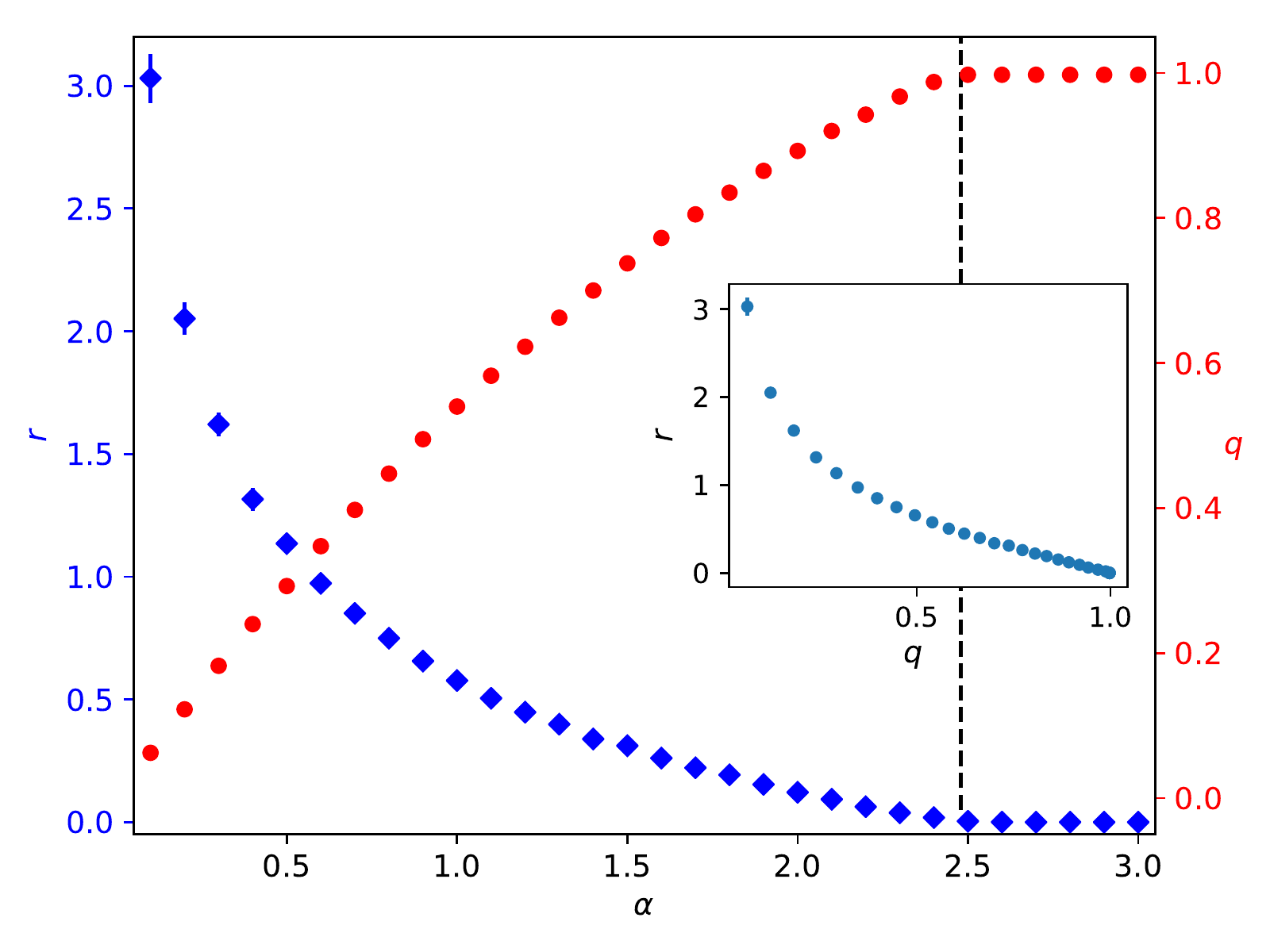}
    \caption{Minimal norm $r=\text{min}_{\mathbf{x}\geq \mathbf{0}} |\Hat{a}^T\mathbf{x}-\mathbf{b}|$ (blue, diamonds) and overlap parameter $q$ (red, dots) as functions of $\alpha$. The results for $q$ where obtained by extremalization of Eq.~(\ref{Sres}). For each data point of the minimal norm we drew 50 realizations of Eq.~(\ref{Equ: Linear Equation System}) and determined the minimal norm by using a least-squares solver. The dotted line indicates  $\alpha_c$. The inset shows a plot of the minimal norm over $q$ revealing a monotonous dependency between the two quantities. Parameters: $\zeta=3$ and $N=600$.}
    \label{fig:Figure 3}
\end{figure}

The Farkas Lemma connects the two dual problems \eqref{eq:nqw} and \eqref{eq:y}: whenever there is a solution to one there is none to the other. Therefore the threshold values $\alpha_c$ for both problems coincide. However, from the solution space analysis described in section~\ref{sec3} more details may be extracted. For $\alpha<\alpha_c$, e.g., there are many different solutions to problem \eqref{eq:y}. The order parameter $q$ which can be determined from the numerical extremalization in \eqref{Sres} is then strictly smaller than one and  provides a measure of the variability between different solutions to \eqref{eq:y}. At the same time the dual problem~\eqref{eq:nqw} is unsolvable as demonstrated by a non-zero residual norm 
\begin{equation}
 r:=\text{min}_{\mathbf{x}\geq \mathbf{0}} |\Hat{a}^T\mathbf{x}-\mathbf{b}|.
\end{equation} 
This residual norm in some sense quantifies how far away problem~\eqref{eq:nqw} is from its threshold of solvability. It is tempting to speculate about a connection between $q$ and $r$. If $q$ is rather small there are many different solutions to \eqref{eq:y}. Accordingly, problem~\eqref{eq:nqw} should be 'far from being solvable' corresponding to a large value of $r$. For $q$ near to one, on the other hand, there is not much freedom left for different solutions of \eqref{eq:nqw} and small values of $r$ are likely. We have tested this conjecture with the help of numerical simulations. The results shown in Fig.~\ref{fig:Figure 3}  indeed suggest a monotonous relation between $q$ and $r$. 

Equivalent considerations apply to the region $\alpha>\alpha_c$. Here \eqref{eq:y} is unsolvable and its solution space is empty. By a slight generalization of the techniques used in section~\ref{sec3} it is possible to determine the minimal number of violated inequalities in \eqref{eq:y} \cite{GaDe}. It is an open question  whether this number may be related to the variability between different solutions of \eqref{eq:nqw} for $\alpha>\alpha_c$.

\section{Conclusion}\label{sec7}

In the present paper we investigated under which conditions a large set of $\alpha N$ random linear equations for $N$ unknowns typically possesses a non-negative solution. This is a rather general question with relevance for a number of different problems in statistical mechanics. It is also an active area of research in mathematics on its own \cite{donoho2005sparse,wang2011unique}. 

With the help of Farkas' lemma we mapped the problem onto the determination of the typical size of a random volume in high-dimensional space which could be characterized analytically using methods from the statistical physics of disordered systems. Our main result is a sharp transition in the limit $N\to \infty$ that separates a phase where the linear system typically has no non-negative solution to one in which such a solution can be found with probability one. The analytical result for the transition line agrees very well with the outcome of numerical simulations. Special cases of the transition have been discussed previously in the literature in connection with the storage capacity of a large perceptron and with a random variant of MacArthurs resource competition model.  

In order to keep the calculations simple we assumed Gaussian distributions for the random parameters. For the elements of the coefficient matrix this assumption is not crucial. As in related mean-field models any distribution of these elements with finite moments would give similar results depending only on their first two cumulants. This is, e.g., exemplified by the application to MacArthurs model. The statistical properties of the inhomogeneity vector are more subtle and similar generalizations to different distributions deserve further investigation. 

The dual problem related to the typical size of a random volume in $N$-space can also be studied for values of $\alpha$ different from the threshold $\alpha_c$. It were rather interesting if this knowledge could be used
to quantitatively characterize the original problem away from criticality, i.e. deeply in the solvable or unsolvable phase. We presented preliminary numerical evidence that this may be possible but more work is needed to substantiate this connection.

\vspace{.5cm}

Acknowledgement:
Financial support from the German Science Foundation under project EN 278/10-1 is gratefully acknowledged. A.E. is deeply grateful for numerous inspiring discussions and genial conversations with Chris van den Broeck he enjoyed over many years.



\bibliographystyle{elsarticle-num}

\bibliography{sample.bib}

\end{document}